\documentclass[
  ,draft            
  ,numberedheadings 
  ]
  {aipproc}

\layoutstyle{6x9}
\usepackage{amsfonts}
\usepackage{amssymb}
\usepackage{amsmath}

\def\p{\phi}

\def\a{\alpha}
\def\b{\beta}

\def\D{\Delta}

\def\t{\theta}
\def\T{\Theta}

\def\pd{\partial}

\def\m{\mu}
\def\r{\rho}
\def\t{\theta}

\def\bra{{\langle}}
\def\ket{{\rangle}}

\def\l{\lambda}

\def\cI{{\cal{I}}}

\def\cA{{\cal{A}}}

\def\cI{{\cal{I}}}

\def\t{\theta}
\def\T{\Theta}
\def\bn{\begin{eqnarray*}}
\def\en{\end{eqnarray*}}
\def\be{\begin{equation}}
\def\ee{\end{equation}}
\def\beq{\begin{eqnarray}}
\def\eeq{\end{eqnarray}}

\begin{document}

\title[Max-Born symposium, ~~~~~~~~T R Govindarajan]
{Spontaneous breaking of symmetry in Moyal spacetime with twisted Poincar\'e
symmetry}


\author{T R Govindarajan}{
address={The Institute of Mathematical Sciences, C I T Campus, Taramani, Chennai 600 113,
India, trg@imsc.res.in},
}
%
%

\begin{abstract}
After briefly reviewing the gauge symmetry in Moyal spacetimes,
we analyse  aspects of symmetry breaking 
within a quantisation program  preserving  the twisted Poincar\'e symmetry.
We develop the LSZ approach for Moyal spacetimes and
derive a mapping for scattering amplitudes on these
spacetimes from the corresponding ones on the commutative spacetime.
This map applies in the presence of spontaneous breakdown of symmetries as well.
We also derive Goldstone's theorem on Moyal spacetime. The formalism developed
here can be directly applied to the twisted standard model
\footnote{Work done in collaboration with A P Balachandran and Sachin Vaidya}.
%
%
\end{abstract}

\maketitle


\section{Introduction}
It has beeen pointed out by many in this Symposium the  
role of  noncommutative geometries when we incorporate 
quantum gravity\cite{sergio}
Noncommutative geometry is expected to play a role in near horizon geometry
of a blackhole\cite{trgnpb} as well as near big bang singularity and cosmological constant\cite{steinacker}. 
The need to go beyond the conventional notions of geometry 
was anticipated by non other than Riemann himself, given 
lack of knowledge of physics at infintesimal length scales.
As pointed out by Riemann ``the metric relations of space in the infinitely small do
not conform to hypotheses of geometry; and we ought in fact to suppose it,
if we can thereby obtain a simpler explanation of phenomena....''\cite{clifford}.
Fuzzy physics as provided by the coadjoint orbits of Lie groups \cite{book,fermiondoubling,trg,denjoe}
and $\kappa$ space-time geometry\cite{lukierski,trgkappa} are classic examples used in physics
literature. One can even consider deformations of Lie algebra 
leading to topology change in the commutative limit\cite{trgpramod}. Here 
we consider the  Moyal space-time which is the most popular example for noncommutative geometry
and look at the way it can 
change our expectations in fundamental physics. 

Moyal space-times are defined by the following algebra:
\be
\left[~\hat{x}_\m,\hat{x}_\nu~\right]~=~i\theta_{\m\nu} {\mathbb I}
\ee
It is well known we can obtain the same through the star product rule
in the algebra of functions on $R^4$. 
\be 
f~*~g~=~m_\t(f \otimes g)~=~m_0(F_\t(f\otimes g))
\ee
where
$F_\t~=~e^{-\frac{i}{2}(-i\pd_\m)\T^{\m\nu}\otimes (-i\pd_\nu)}$.
In commutative spacetime we have pointwise multiplication.
\subsection{Drinfeld twist}
As pointed out by Balachandran\cite{sergio,chaichian,baltrg} we have to 
find a twisted coproduct $\D_\t$ compatible with the multiplication map:
\be
m\left[ \left( \r\otimes \r\right) \Delta(g)\left(a
\otimes b\right) \right] ~=~\r(g)~m(a \otimes b) 
\ee
where $a,b \in \cA_\t(R^4)$.
The above can be shown as commutative diagram(Fig 1).
\begin{figure}
  \includegraphics[height=.1\textheight]{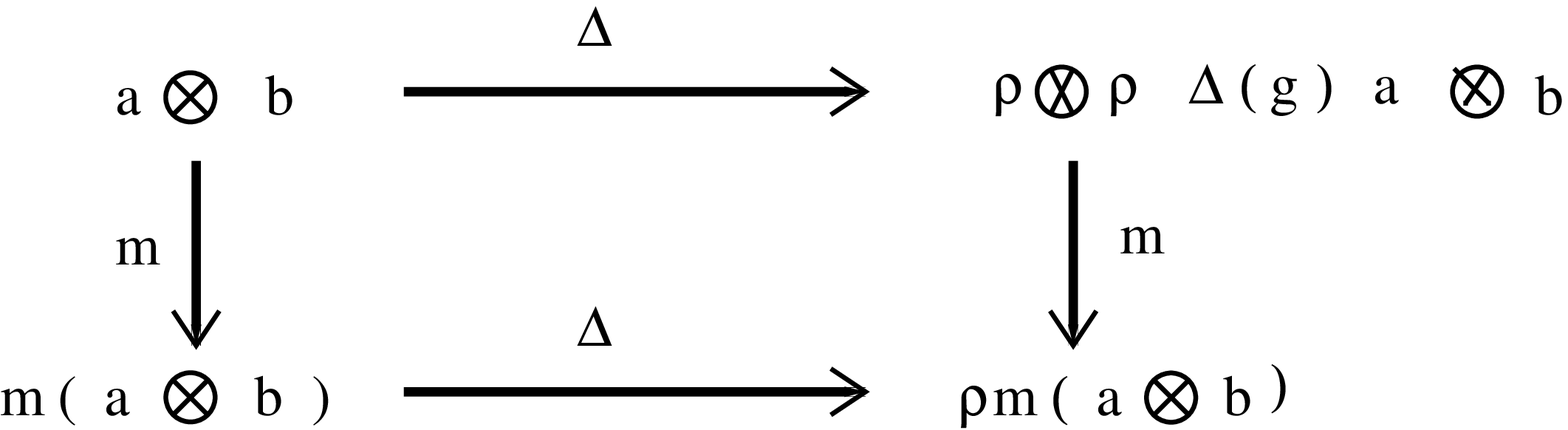}
  \caption{commutative diagram}
\end{figure}

Indeed such a twisted coproduct for Moyal space is:
\be
\D_\t(g)~=~{\hat{F}}_\t^{-1}(g\otimes g){\hat{F}}_\t
\ee
where ${\hat{F}}_\t~=~e^{-\frac{1}{2}~P_\m \otimes \t^{\m\nu}~
P_\nu}$, $P_\m$ is the generator of translations.
It is easy to check that the coproduct is compatible with the
multiplication map.
\be
 m_\t(\r \otimes \r)\D_\t(g)(a \otimes b)~=~ m_0\left[F_\t(F_\t^{-1}
\r(g)\otimes \r(g)~F_\t)a \otimes b\right]
\ee
\subsection{Twisted Poincar\'e invariance and statistics}
For $\theta^{\mu \nu}=0$ statistics is imposed on
the two-particle sector by working with the symmetrized or anti-symmetrized tensor product
${\cal A}_0 ({\mathbb R}^4) \otimes_{s,a} {\cal A}_0 ({\mathbb R}^4)$. 
It has for example
\be
v \otimes_{s,a} w = \frac{1}{2}[v \otimes w \pm  w\otimes v], \quad v,w \in
{\cal A}_0 ({\mathbb R}^4).
\ee
But the twisted coproduct does not preserve (anti)symmetrization:
\be
\Delta_\theta (v \otimes_{s,a} w) \notin {\cal A}_0 ({\mathbb R}^4)
\otimes_{s,a} {\cal A}_0 ({\mathbb R}^4)
\ee
We are forced to twist statistics. To achieve this consider 
$\tau_0$ to be the flip map:
\be
\tau_0 (v \otimes w) = w \otimes v.
\ee
Then
\be
\tau_\theta :=F_\theta^{-1} \tau_0 F_\theta = F_\theta^{-2} \tau_0
\ee
commutes with $\Delta_\theta $. The product ${\cal A}_\theta ({\mathbb R}^4)
\otimes_{s_\theta,a_\theta} {\cal A}_\theta ({\mathbb R}^4)$ with twisted
(anti-) symmetrization is :
\be
v \otimes_{s_\theta,a_\theta} w = \frac{1}{2}[I \pm  \tau_\theta](v
\otimes w)
\ee
Imposing this change on the quantum fields leads to the commutation relations\cite{baltrg}:
\begin{eqnarray}
a(p) a(q) &=& e^{i p \wedge q} a(q) a(p), {\rm and} \nonumber \\
a(p) a^\dagger (q) &=& e^{-i p \wedge q} a^\dagger (q) a(p) + 2p_0
\delta^{3}(p-q).
\end{eqnarray}
\section{Diffeomorphism and gauge invariance in Moyal space-times}
To define diffeomorphisms and gauge symmetries in this framework is 
very complicated. We will adopt a novel way\cite{baltrgvaidya}.
Consider $x_\m^c ~=~
\frac{1}{2}(x_\m^L~+~x_\m^R)$ where $x_\m^L~\a~=~x_\m*\a$
and $x_\m^R~\a~=~\a *x_\m$.
It is easy to see
\be
\left[x_\m^c, x_\nu^c\right]~=~0.
\ee
This simply means $x_\m^c$ form a basis for commutative algebra
${\cal A}_0({\mathbb R}^4)$.  One can define Poincar\'e group of generators using
$x_\m^c$ as
\be
M_{\m\nu}~=~x_\m^c~p_\nu~-~~x_\nu^c~p_\m~, p_\m~=~-i\pd_\m
\label{Mmunu}
\ee
We get modified Leibnitz rule:
\beq
M_{\m\nu}(\a * \b)&=&M_{\m\nu}\a*\b~+~\a*M_{\m\nu}\b~ \nonumber \\
&-&\frac{1}{2}[
(p.\t)_\m \a*p_\nu\b-(p_\nu\a*(p.\t)_\mu\b-\m\leftrightarrow\nu]
\eeq
This is exactly the same as what we get from twisted coproduct $\D_\t$.
In the above $M_{\m\nu}$ in Eq.(\ref{Mmunu}) is a particular vector field.
This can be extended to general vector fields
$v~=~v^\m(x^c)\pd_\m$. These generate the diffeomorphisms on the Moyal spacetime.
If we assume the framefields $e_\mu^a$ are dependent
only on $x^c$ then pure gravity without matter can be treated as in commutative spacetimes.
Gauge fields $A_\lambda$ transform as one-forms under
diffeomorphisms for $\theta^{\mu \nu}=0$. For $\theta^{\mu \nu}
\neq 0$, the vector fields $v^\mu$ generating diffeomorphisms
depend on ${x}^c$. If a diffeomorphism acts on $A_\lambda$
in a conventional way with $A_\lambda$ and $\delta A_\lambda$ are to depend
on just one combination of noncommutative coordinates, then
$A_\lambda$ can depend only on ${x}^c$. In the standard approach to 
gauge symmetry in noncommutative geometry  the 
covariant derivatives act with the $*$ -product and this imposes severe constraints. It is possible to
have  only particular representations of $U(N)$ gauge groups in the theory.  
We cannot impose standard model group consistently. There is no such limitation now
in our novel way of introducing gauge transformations.
This approach was inspired by quantum Hall effect, where guiding center coordinates 
act as a model of Moyal space-times\cite{balkumar}.
Here the algebra of observables is ${\cal A}_\theta
({\mathbb R}^2) \otimes {\cal A}_\theta ({\mathbb R}^2)$.
Here too covariant derivatives of the $U(1)$ electromagnetism do act in the way we have
described above and
not with a $*$ product. But the twisted coproduct on the ``global'' group ${\cal G}$ is,
\be
\Delta_\theta (g(x^c) = F_\theta^{-1} [g(x^c) \otimes
g(x^c)] F_\theta,
\ee
and is compatible with the $*$ -multiplication.
\subsection{Dressing transformation}
The creation/annihilation operators {$a(p), a^\dagger(p)$} 
for the twisted fields can be realized in terms of untwisted Fock space
operators {$c(p),c^\dagger(p)$} by the
``dressing transformation" \cite{grosse}
\beq
a(p) &=& c(p) e^{-\frac{i}{2} p \wedge P}, \quad a^\dagger (p) = c^\dagger (q)
e^{\frac{i}{2} p \wedge P}, {\rm where} \nonumber \\
P_\mu &=& \int d\mu(q) q_\mu [a^\dagger(q) a(q)] =
{\rm total \, momentum\, operator} 
\eeq
where $d \mu(p) \equiv \frac{d^3p}{2p_0}, p_0=\sqrt{\vec{p}^2 + m^2}$ is the Lorentz invariant measure.
Then {$\phi(x)$} may be written in terms of commutative
fields {$\phi^c$} as
\be
\phi(x) = \phi^c e^{\frac{1}{2}\overleftarrow{\partial} \wedge P}(x)\,.
\ee
If $\phi_1,\phi_2,\cdots \phi_n$ are quantum fields and 
$\phi_i(x) = \phi_i^c e^{\frac{1}{2}\overleftarrow{\partial} \wedge P}(x)$,
then
\be
(\phi_1* \phi_2*\cdots \phi_n)(x) = (\phi_1^c
\phi_2^c \cdots \phi_n^c)e^{\frac{1}{2}\overleftarrow{\partial} \wedge P}(x)
\ee
For example the interaction Hamiltonian density is:
\be
{\cal H}_{I \theta} = {\cal H}_{I 0} \; e^{\frac{1}{2} \overleftarrow{\partial}
\wedge P}
\ee
The covariant derivative should transport consistently with the
statistics as well as gauge transformations and hence the natural choice is:
\be
D_\mu \phi = ((D_\mu)^c \phi^c)e^{\frac{1}{2}
\overleftarrow{\partial} \wedge P}
\ee
where
$(D_\mu)_c \equiv \partial_\mu + (A_\mu)_c$
and $(A_\mu)_c$ is the commutative gauge field, a function only of $x^c$.
It is easy to check:
\be
[D_{\mu}, D_{\nu}] \varphi~=~
\Big([D^{c}_{\mu}, D^{c}_{\nu}]\varphi^{c}\Big)e^{\frac{1}{2}
\overleftarrow{\partial} \wedge P}
~=~\Big(F_{\mu \nu}^{c}\varphi^{c}\Big)
e^{\frac{1}{2}\overleftarrow{\partial} \wedge P}.
\ee
We can also write
\be
D_{\mu}\varphi~=~\Big(D_{\mu}^{c} e^{\frac{1}{2}\overleftarrow{\partial} \wedge
P}\Big)\star \Big(\varphi^{c}e^{\frac{1}{2}\overleftarrow{\partial} \wedge
P}\Big).
\ee
As $F_{\mu \nu}^{c}$ is the
standard $\theta^{\mu \nu}=0$ curvature,
gauge field is that of commutative space-time and transforms covariantly under
gauge transformations. We can use it to construct the Hamiltonian.
\subsection{Gauge theory on Moyal space-time}
Having set up the necessary formalism, we can now write down the interaction Hamiltonian density 
for pure gauge fields as:
\bn
{\cal H}_{I \theta}^{^G} = {\cal H}_{I 0}^{^G}.
\en
But when both matter and gauge fields are present, the interaction Hamiltonian density is:
\be
{\cal H}_{I \theta} = {\cal H}^{^{M, G}}_{I \theta}+{\cal H}^{^G}_{I \theta},
\ee
where
\be
{\cal H}^{^{M, G}}_{I \theta}~=~{\cal H}^{^{M, G}}_{I 0} \; e^{\frac{1}{2}
\overleftarrow{\partial} \wedge P}
\ee
In $QED_\t$, we have ${\cal H}^{^G}_{I \theta}=0$.
\be
S^{QED}_{\theta}=S^{QED}_{0}.
\ee
On the otherhand in  $QCD_\t$, we have
${\cal H}^{SU(3)}_{I\theta}={\cal H}^{SU(3)}_{I0} \neq 0$,
so that
\be
S^{M,SU(3)}_{\theta} \neq S^{M,SU(3)}_{0}.
\ee
Lastly we can also look for Standard model$_\t$
after discussing spontaneous symmetry breakdown. But first  we will develope
Lehmann, Symanzik, Zimmerman (LSZ) formalism suitable for 
Moyal spacetime which will facilitate computations in scattering theory. 
\section{LSZ on Moyal spacetime}
In standard scattering theory, the Hamiltonian $H$ is split into a
``free'' Hamiltonian $H_0$ and an ``interaction'' piece $H_I$:
$H_0$ is used to define the states in the infinite past and
future. Then the states at $t=0$ which in the infinite past (future)
become states by evolving $H_0$ as the in(out) states.
\beq
e^{-i H T_\pm}|\psi, {\rm {in(out)}}\rangle &\stackrel{T_\pm \rightarrow \pm \infty}
\longrightarrow e^{-i H_0 T_\pm} |\psi, {\rm F}\rangle, \quad {\rm F}
\equiv {\rm free}
\eeq
Hence
\be
|\psi, {\rm in(out)}\rangle = \Omega_\pm |\psi,{\rm F}\rangle,
\Omega_\pm \equiv e^{i H T_\mp} e^{-i H_0 T_\mp},~~{\rm as}~~
T_\pm \rightarrow \pm \infty,
\ee
Here $\Omega_\pm$ are the Moller operators.
We have:
\be
|\psi,{\rm out} \rangle = \Omega_- \Omega_+ |\psi,{\rm in}\rangle
\ee
If the incoming(outgoing) state is $|k_1, k_2, \cdots k_N, {\rm F}\rangle$, it
follows that
\be
|k_1, k_2, \cdots k_N, {\rm in(out)}\rangle = \Omega_\pm |k_1, k_2, \cdots
k_N, {\rm F}\rangle
\ee
has eigenvalue $\sum k_{i0}$ for the total Hamiltonian $H$.
The scattering amplitude is:
\be
\langle \psi,{\rm out} | \psi,{\rm in}\rangle = \langle \psi,{\rm
in}|\Omega_+ \Omega_-^\dagger |\psi,{\rm in}\rangle
\ee
The LSZ {$S$}-matrix is
\be
S = \Omega_+ \Omega_-^\dagger, \quad |\psi,{\rm out} \rangle =
S^\dagger |\psi,{\rm in}\rangle
\ee
Between the ``free'' states, the {$S$}-operator is different:
\be
\langle \psi,{\rm out} | \psi,{\rm in}\rangle = \langle \psi,{\rm
F}|\Omega_-^\dagger  \Omega_+ |\psi,{\rm F}\rangle
\ee
The LSZ formalism works exclusively with in- and out-states, as Haag's
theorem shows that {$\Omega_\pm$} do not exist for quantum field
theories.
The operators {$a_k^{\rm in
(out)\dagger}, a_k^{\rm in(out)}$} are introduced to create states
$|k_1,k_2,\cdots k_N, {\rm in(out)}\rangle$ from the vacuum. The in-
and out- fields $\phi_{\rm in(out)}$ are then defined using
superposition. They look like free fields, but are not, since for
the total four-momentum $P_\mu$, we have
\be
P_\mu |k_1,k_2,\cdots k_N, {\rm in(out)}\rangle = (\sum_i k_{i \mu})
|k_1,k_2,\cdots k_N, {\rm in(out)}\rangle .
\ee
We also assume:
\begin{enumerate}
\item
The vacuum and single particle states are unique. 
\item
There is only one vacuum {$|0\rangle$, $\langle 0 | 0\rangle =1$},
and
\be
S|0\rangle = |0\rangle
\ee
\item There exists an interpolating field {$\phi$}
between in- and out- states in weak topology:
\be
\phi - \phi_{\rm in, out} \rightarrow 0 \quad {\rm as} \quad \tau
\rightarrow \pm \infty
\ee
\end{enumerate}
Then LSZ show that
\be
\langle k'_1,\cdots k'_N, {\rm out}|k_1,\cdots k_N {\rm
in}\rangle = \cI~~G_N(x_1,x_1',\cdots,x_N,x_N')
\label{ncLSZ}
\ee
where
\be
\cI~=~\int\prod d^4x'_i\prod d^4x_j~e^{i(k_i'\cdot x_i'~-~k_j\cdot x_j)}
i(\pd_i'^2+m^2)\cdot i(\pd_j^2+m^2)
\ee
and
\be
G_N \equiv \langle 0|T(\p(x_1)\p(x_1')\cdots \p(x_N)\p(x_N'))|0\rangle
\label{scatter}
\ee
We have argued that the noncommutative field theory comes from the
commutative one by the replacement:
\be
\phi_\theta = \phi_0 e^{\frac{1}{2} \overleftarrow{\partial} \wedge
P}.
\ee
This consistently twists the in- and out- fields:
\be
\phi_\theta^{\rm in,out} = \phi_{\rm in, out}
e^{\frac{1}{2}\overleftarrow{\partial} \wedge P},~~
\phi_\theta \rightarrow \phi_\theta^{\rm in,out} \quad {\rm as} \quad
t \rightarrow \pm \infty
\ee
LSZ holds for scattering amplitude with $G_N$ changed to $G_N^\theta$:
\begin{eqnarray}
G_{N}^\theta (x_1, \cdots x_{N}) &=& T e^{\frac{i}{2}\sum_{I<J}\partial_{x_I}
\wedge \partial_{x_J}} W_{N}^0 (x_1, \cdots x_{N}) \nonumber \\
:&=&~T~W_{N}^\theta (x_1, \cdots x_{N})
\label{GNtheta}
\end{eqnarray}
and $W_{N}^0$ are the standard Wightman functions for untwisted fields:
\begin{equation}
W_{N}^0 (x_1, \cdots x_{N}) = \langle 0 | \phi_0 (x_1) \cdots \phi_0(x_{N})
|0\rangle.
\end{equation}
It is important that because of translational invariance, the
$W_{N}$ (and hence the $G_{N}$) depend only on coordinate
differences.
For simplicity, we have included only matter fields, and that too of one type only, in
(\ref{scatter}). Gauge fields can also be included, but they are not acted on by the twist
exponential in (\ref{GNtheta}).
\subsection{Gell-Mann-Low formula for Moyal spacetime}
We assume Heisenberg fields $\phi$ obey the same canonical algebra
as free fields $\phi_F$ at $t=0$.
The interaction representation Hamiltonian is
\be
H_I(t) = e^{it H_0} H_I(0) e^{-i t H_0}, \quad H_I(0) = H_I
\ee
The time evolution operator is:
\be
U(t_1,t_2) = T \exp \left( -i \int_{t_1}^{t_2} dt H_I (t) \right),
\ee
Then Gell-Mann and Low (G L) formula show:
\be
G_N(x_1,x_2,\cdot x_N)= \frac{\langle 0,{\rm F}|T \left(\phi_F(x_1) \cdots
\phi_F(x_N) e^{i\int d^4 x {\cal L}_I (x)}\right) |0,{\rm
F}\rangle}{\langle 0,{\rm F}| e^{i\int d^4 x {\cal L}_I (x)}|0,{\rm
F}\rangle}
\ee
G L formula applies to $G_N^0$. We can rewrite the result in terms
of the Wightman functions $W_N (x_1, \cdots x_N)$:
\be
W_N (x_1, \cdots x_N) = \langle 0 | \phi (x_1) \cdots \phi(x_N)
|0\rangle
\ee
Then,
\be
G_N^\theta (x_1, \cdots x_N) = T e^{\frac{i}{2}\sum_{I<J}\partial_{x_I}
\otimes \partial_{x_J}} W_N (x_1, \cdots x_N)
\ee
This results in  shifting of each $x_I^0$ to
\be
x_I^0 + \delta x_I^0, \quad \delta x_I^0 = \delta x_I^0 (k_1, \cdots k_N).
\ee
where the $\delta x_I^0$ actually depend on the ordering on $x_I^0$.

\noindent We emphasize couple of important observations.
\begin{enumerate}
\item Firstly, (\ref{ncLSZ}) involves only the $\theta^{\mu\nu}=0$
fields in $W_N^0$. So it can be used to map any commutative theory to
noncommutative one, including the standard model. But special
care is needed to treat gauge fields.  Gauge fields are {\em{not}}
twisted unlike matter fields. As explained elsewhere, this means 
the Yang-Mills tensor is not twisted,
$F^{\mu\nu}_\theta~=~F^{\mu\nu}_0$. But covariant derivatives of
matter fields $\phi_\theta$ are twisted: $(D_\mu\phi)_\theta~=~
(D_\mu\phi)_0~e^{\frac{1}{2}{\overleftarrow{\partial}}\wedge
P}$. where $(D_\mu\phi)_0$ is the untwisted covariant derivative of
the untwisted $\phi_0$.  Thus in correlators $W_N^\theta$, we must use
$(D_\mu\phi)_\theta$ for matter fields, $F^{\mu\nu}_0$ for Yang-Mills
tensor.
\item There are ambiguities in formulating scattering theory.
At this moment, lacking a rigorous scattering
theory, we do not know the correct answer.  In this
connection, we must mention the important work of Buchholz and
Summers\cite{buchholz} which rigorously develops the wedge
localisation ideas of Grosse and Lechner\cite{lechner} to establish a scattering
theory for two incoming and two outgoing particles.
\end{enumerate}
For calculating (\ref{ncLSZ}), we need a formalism for doing
perturbation theory to compute Wightman functions. Once we have
that, we can calculate the time-ordered product by writing it in terms
of Wightman functions and twist factors. The details of these can be found in 
our paper \cite{baltrgvaidya}.
\subsection{Golstones' theorem on Moyal spacetime}
We will now take up the question of the spectrum in Moyal spacetimes. In 
answering this question spontaneous breakdown of symmetry and Goldstones'
theorem plays an important role. The noncommutative
currents {$J^{a,\mu}_\theta (x) = J^{a,\mu}_0(x)e^{\frac{1}{2}
\overleftarrow{\partial} \wedge P}$} are conserved:
\be
\partial_\mu J^{a,\mu}_\theta (x) = \partial_\mu J^{a,\mu}_0(x)e^{\frac{1}{2}
\overleftarrow{\partial} \wedge P} = (\partial_\mu
J^{a,\mu}_0(x))e^{\frac{1}{2} \overleftarrow{\partial} \wedge P} = 0.
\ee
The vacuum expectation value of the currents
{$J^{a,\mu}_\theta(y)$} and the quantum field {$\phi_{i,\theta}(x)$}:
\be
\langle 0| [e^{\frac{1}{2}\overrightarrow{\partial_y} \wedge P} J^{a,\mu}_0(y),
\phi_{i,0}(x)e^{\frac{1}{2} \overleftarrow{\partial_x} \wedge P}] |0\rangle
= \langle 0|[J^{a,\mu}_0(y), \phi_{i,0}(x)]|0\rangle
\ee
This commutator is the same as the one for the corresponding
commutative case. Using spectral density and twisted Lorentz invariance we  can infer
the existence of massless bosons.
\section{Higgs{$_\t$} mechanism}
Consider a set of scalar fields coupled to a gauge field
with a local symmetry group {$G$}. This dynamics is described by:
\be
{\cal L} = {\rm Tr} \left(-\frac{1}{4} F_{\mu\nu}^2 + |D_\mu
\p|^2 - V(\p) \right)
\ee
The  Higgs potential 
\beq
V(\p)&=&\l(\p^\dagger*\p~-~a^2)^2_* \\
&=&\l(\p^\dagger_c\p_c~-~a^2)~e^{\frac{1}{2}
\overleftarrow{\partial} \wedge P}
\eeq
We assume the breaking {$G\longrightarrow H$}. In vacuum
\be
\bra \p_c \ket =\p^0,~~\p^{0\dagger}~\p^0~=~a^2, ~~h~\p^0=\p^0,
h~\in~H
\ee
\subsection{Mass of the gauge boson}
The vacuum manifold is:
\be
\p~=~g~\p^0,~~g~\in~G, ~and~~~ (gh)~\p^0~=g~\p^0
\ee
The gauge field acquires mass and is given by the term:
\be
M~=~(D_\m\p)^\dagger*(D^\m\p)~=~[(D_\m^c\p_c)^\dagger({D^\m}^c\p_c)]
e^{\frac{1}{2}
\overleftarrow{\partial} \wedge P}
\ee
If $V(\a),S(i)$ are basis of
orthonormal generators of Lie algebra
$\bf G$ of $G$, then:
\be
V(\a)\p^0~=~0
\ee
If a gauge transformation is performed from
$A_\m^c \rightarrow B_\m^c$
where {$B_\m^c~=~g^\dagger~D_\m^c~g$}, then,
\be
M~=~\p^c{^\dagger}_\a(B_\m^c{^\dagger} B^{\m c})_{\a\b}\p^c_\b
\ee
As usual we write:
\be
B_\m^c~=~B_\m^c{^\a}V_\a~+~~B_\m^c{^i}S_i~
\ee
Then we get:
\be
M~=~(D_\m^c\p^c)^\dagger (D^\m{^c}\p^c)~=~\p^0{^\dagger}S_iB_\m^iB^\m{^j}S_j
\p^0~+~\cdots
\ee
This shows gauge fields in the direction of {$V_\a$}
do not acquire mass
and only those in the direction of {$S_i$} do.
{$B_\m^c$} is the gauge transformation of
{$D_\m^c$}. This preserves
the  pure gauge Hamiltonian {$H_I{_\t}~=~H_I{_0}$}.
After gauge fixing the Hamiltonian with the mass term is:
\be
H_0~=~\int \{\pd\wedge B^c)^2~+~(\pd_0B^i-\pd^iB_0)^2 +\cdots +M\}
\ee
The Hamiltonian $P_0~=~H$ and the spatial translation generator for
the twisted standard model are the same as for the case
$\theta^{\mu\nu}~=~0$ except for the twist factors. 
What is changed in addition in our LSZ approach are the in
and out fields which are twisted as discussed. Hence the scattering
calculations can be based on appropriately modified Wightman functions.
Detailed calculations for specific processes will be considered in our future
papers.

\begin{theacknowledgments}
This work is based on the joint work with A P Balchandran(APB)  and Sachin Vaidya(SV).
It was supported by the Department of Science and Techonlogy, India  program
CPSTIO. I thank APB and SV for discussions. 
\end{theacknowledgments}

\end{document}